\documentclass[11pt,twoside]{article}
\usepackage{newpasp}
\usepackage{epsf}

\index{Stocke, J. T.}

\begin{document}

\title{Hidden BL Lacertae Objects Near and Far}
\author{John T. Stocke}
\affil{Center for Astrophysics \& Space Astronomy, and Dept. of
Astrophysical
\& Planetary Sciences, University of Colorado, Boulder, CO 80309-0389}

\begin{abstract}
I describe two difficulties with current research into the cosmic
evolution of BL Lacertae Objects: (1). Possible sample incompleteness
due to unrecognized (i.e., ``hidden'') BL Lacs; and (2). The absence of
a viable physical model of the evolution, to which current and
future observations
can be compared.
\end{abstract}


\section{Introduction}

The unusual evolutionary trend seen in the number and/or
luminosity of BL Lac Objects (a.k.a. ``negative'' evolution) 
has been the subject of many recent papers
and of several talks at this conference. But two things remain
frustrating (for me at least) involving these evolutionary studies:

1. Unlike quasars (or other AGN classes for that matter), the
identification of a complete BL Lac sample is fraught with difficulty and
uncertainty because BL Lac Objects are defined in ``negative'' terms 
(e.g., they
lack the strong emission lines of quasars). Other defining qualities may
also eliminate true BL Lacs from samples under study, making them
incomplete in a systematic way. 

2. But like the quasars (and other AGN classes as well), there is no
physical model to which the evolutionary trends are compared; i.e.,
functions are fit and argued over (e.g., pure luminosity evolution vs.
luminosity-dependent density evolution) but these are algebraic
functions without much physical basis.

In this brief account I will address both of these issues with respect
to current samples as well as what has been or will be seen in deep
{\bf CHANDRA} images.

\section{Incompleteness in Current Samples}

In this Section I will describe three ways in which true BL Lacs can be
``hidden'' in X-ray and radio-selected samples, potentially causing
incompleteness:

1. Despite many X-ray selected BL Lacs (XBLs) having spectral energy
distributions (SEDs) which peak in the optical/UV (so-called ``high energy
peaked BL Lacs'': HBLs), their optical spectra often have a significant
contribution from host galaxy starlight. Browne \& March\~a (1993) and
March\~a \& Browne (1995) pointed out that such objects could be
confused with FR-1 radio galaxies in clusters so that the X-ray emission
would be assigned incorrectly 
to the cluster. Thus, some XBLs, particularly those
with modest optical non-thermal luminosity, could be
misidentified as clusters of galaxies.

2. Investigators of radio surveys used to 
identify radio selected BL Lac (RBL) samples
have set a radio spectral index cutoff to eliminate the large number
of radio galaxies in these samples (e.g., the 1 Jy sample uses
$\alpha \geq$-0.5; Stickel et al. 1991). However, some XBLs have radio
spectral indices steeper than this, suggesting that a flat radio
spectral index is not required for all BL Lacs. By this spectral index
fiat RBLs would be
misidentified as radio galaxies.
 
3. Both RBLs and XBLs could be ``hiding'' in rich clusters of galaxies
where the radio emission would be ascribed to a brightest cluster
galaxy (BCGs like NGC 1275=3C84=Perseus A) and the X-ray emission ascribed
to dense, cooling gas (i.e., a ``cooling flow'').   
These BL Lacs would be misidentifed as FR1 radio galaxies in
``cooling flow'' clusters.

\subsection{Hidden XBLs}

A large gallery of XBL spectra can be found in the recent compilation
paper of the EMSS XBL sample by Rector et al. (2000). These spectra show
that, unlike RBLs which often have weak (low equivalent width and low
luminosity) emission lines in their optical spectra, XBLs only very
rarely have emission lines but often have significant contributions from
host galaxy starlight. In my first examination of XBLs (Stocke et al.
1991), I suggested a
BL Lac classification criteria to separate BL Lacs from FR 1
radio galaxies using the strength of  
CaII H \& K break at $\approx$4000\AA\, rest. 
A Ca II ``break strength'' of $\leq$25\%
was required for BL Lacs, 
whereas 50\% is typical of current epoch ellipticals (e.g., Dressler \&
Shectman 1987). It is important to understand that, like 
the radio spectral index limit, 
the Ca II break limit is arbitrary, and so Maria March\~a in her thesis
correctly pointed out that this criteria could prevent us from correctly
identifying low luminosity BL Lac Objects in X-ray selected samples
(Browne \& March\~a
1993; March\~a \& Browne 1995). Other, more imaginative, but still
arbitrary, criteria have been proposed since then (e.g., March\~a et al.
1996). These investigations led us (Rector, Stocke \& Perlman 1999) to
use the {\bf ROSAT} HRI to image some poor clusters in the EMSS which
might actually be BL Lac Objects al\'a March\~a. She was correct! 
We found that some of these EMSS sources were pointlike at radio
galaxy locations: new EMSS BL Lacs misidentified as clusters and thus
``hiding'' in these clusters. 
However, some of these new BL Lacs had optical spectra whose Ca II
breaks violated both the old Stocke et al. (1991) and the newer March\~a et al.
(1996) criteria.  
Two are nearly indistinguishable from FR 1 radio galaxies or normal
elliptical galaxies in their Ca II breaks, and yet they are still quite luminous
X-ray point sources (L$_x \approx$10$^{44}$ ergs s$^{-1}$). Optical
spectra of these new BL Lacs can be found in Rector, Stocke \& Perlman
(1999). An important result, 
however, is that even with these new BL Lacs included, the $<V/V_{max}>$
is still significantly less than 0.5, especially for extreme XBLs which
have $<V/V_{max}>$=0.27$\pm$0.08 (Rector et al. 2000), solidifying
``negative'' evolution for XBLs.

Are there other ``hidden'' BL Lacs of this sort? I suspect, yes! Here I
mention only two interesting cases: 

(1) The Owen et al. (1996) optical
spectroscopy of rich cluster radio galaxies which failed to find many BL
Lac candidates (3C264 and IC310 in Perseus being two of the four
possibilities). Given the above, 
I suspect that {\bf CHANDRA} X-ray cluster 
imaging will find new
examples of luminous point X-ray sources associated with cluster radio
galaxies without obvious non-thermal continua optically (see Section
2.3).

(2) Numerous investigators have been bewildered to find faint X-ray
sources (some with quite hard X-ray spectra) associated with ``passive''
elliptical galaxies. By ``passive'' these investigators have meant that
they fail to find obvious AGN signatures optically (e.g., non-thermal
continuum + strong emission lines). By way of example, I direct the
reader to the Nature article by Mushotzky, Cowie, Barger \& Arnaud
(2000) on optical identifications in a {\bf CHANDRA} deep field, in which
three relatively bright, ``passive'' ellipticals have been found. The
optical characteristics of these galaxies are quite similar to the
low-luminosity BL Lacs in Rector, Stocke \& Perlman (1999). So, I 
predict that these sources will have SEDs similar to other BL Lac
Objects (i.e., radio detections at sub-mJy levels).
And one only has to redshift the SEDs displayed in Sambruna, 
Maraschi \& Urry (1996) to realize that extremely hard soft X-ray spectra are
possible for BL Lacs without requiring any internal absorption (as has
been proposed for other potential hard-spectrum X-ray background
contributors like obscured Seyferts).

\subsection{Hidden RBLs}

Radio continuum images of XBLs in Perlman \& Stocke (1993) and Rector,
Stocke \& Perlman (1999) show that many have luminous extended
structure, which is likely to be steep in radio spectral index. Single
epoch radio spectral indices have been obtained for only 8 XBLs (Stocke
et al. 1985), for which only half would pass the -0.5 spectral index
cutoff used to define a BL Lac candidate list in the 1 Jy sample
(Stickel et al. 1991). Spectral index observations of the entire EMSS
sample are now being obtained to see how important the arbitrary
spectral index cut imposed upon RBL samples is (Wolter, Rector \&
Stocke, in progress). But, given the current data, I suspect that there
could be many BL Lacs ``hidden'' amongst the large number of radio
galaxies in the 1 Jy. Some of these could have optical spectra similar
to the low-luminosity XBLs in Rector, Stocke \& Perlman (1999), further
complicating their identification (e.g., 3C264). Unfortunately, optical
polarimetry blueward of the Ca break may be the only viable observation
useful in testing this prediction. But, if enough of these new BL Lacs
are found in the 1 Jy, their presence could alter the $<V/V_{max}>$
statistic for that sample.

\subsection{Other Hidden BL Lacs}

A detailed imaging survey of BL Lacs conducted at the
Canada-France-Hawaii 3.6m several years ago (Wurtz et al. 1996, 1997)
presented a confusing result: BL Lacs were not found in the brightest
cluster galaxies (BCGs) or in the richest clusters (Abell richness
classes $>$ 1) in the current epoch. How can this be if the parent
population of BL Lacs are FR1s, which are abundant in nearby rich
clusters (e.g., Owen et al. 1996 and references therein)? A 
possibility, not yet fully explored, but now possible with {\bf CHANDRA}, is
that cluster X-ray emission ``hides'' some BL Lac Objects. This
is especially true in the case of ``cooling flow'' clusters, whose
central X-ray excesses are thought to be due to a dense, cooling ICM. 
However, if some BCGs in these rich clusters are unrecognized BL
Lac Objects, their X-ray emission could contribute to the excess thought
to be a ``cooling flow''. Because the soft X-ray spectra of XBLs are
similar to ICM emission (see e.g., the {\bf Beppo-SAX} spectra 
in Wolter et al. 1998), these XBLs could ``masquerade'' as ``cooling
flows''. But, one might argue that the
optical spectra of BCGs in ``cooling flow'' clusters rules these objects
out as BL Lacs, since strong emission lines are present (particularly
[OII] and H$\alpha$). But these emission lines arise in spatially very
extended gas (tens to hundreds of kpc) and so are {\bf not} the nuclear
emission lines of an AGN. I predict that {\bf CHANDRA ACIS} imaging of
nearby ``cooling flow'' clusters will discover luminous point X-ray sources
associated with many BCGs (BCGs in ``cooling flow'' clusters are very often
radio galaxies!; Burns 1990). The presence of these point sources
will indicate previously unknown BL Lac Objects and will reduce substantially
the required mass inflow for the ``cooling flow''. We have searched the EMSS
clusters for such circumstances using ROSAT HRI images and have found
only one cluster which could contain a BL Lac in its BCG: MS1455.0+2232.
While {\bf CHANDRA} images will test this suggestion, this one ``hidden'' 
BL Lac will not alter the statistics of the EMSS sample.

\section{Models for AGN Evolution ?}

So, despite the substantial concerns about completeness described
above, the EMSS XBL sample has been very thoroughly scrutinized for
``hidden'' BL Lacs and continues to exhibit ``negative'' evolution.
Further, recent work on other AGN samples (e.g., Urry \& Padovani 1995)
find that BL Lac and FR1 evolution can be characterized as L(z)/L(z=0)=
(1+z)$^{\Gamma}$, with $\Gamma$=-4.0, while FR2 samples have $\Gamma$=+3.8.
These results suggest that FR2s evolve (i.e., fade to become) FR1s. Other
indications that this is the case comes from the investigations into
the environment of quasars (Ellingson, Yee \& Green 1991), FR2 radio
galaxies (Harvanek et al. 2000) and FR1 radio galaxies (e.g., Longair \&
Seldner 1979; 
Yee \& Lopez-Cruz 1999). Briefly, from these works: quasars are
found in clusters only at $z\geq$0.45, FR2s are found in clusters only
at $z\geq$0.2 and FR1s are found in clusters at all redshifts out to 0.8
(e.g., Stocke et al. 1999). 

What can we make of these results? Is a 
physical model suggested or is any suggestion premature? 
To me these results suggest two things: 
(1) the
orientation unification advocated by Bartel (1989) is ruled out since
some FR2 radio galaxies are found in clusters at lower redshifts than
any quasars are found in clusters; and (2) an evolutionary scheme by
which quasars fade to become FR2 radio galaxies which fade to become FR1
radio galaxies on an e-folding timescale of 0.9 Gyrs (Harvanek et al.
2000) is quite plausible for cluster AGN. 
After all, what has become of the quasars
which inhabited rich clusters at z$\sim$0.5? This hypothesis is also
consistent with the evolutionary trends for BL Lacs/FR1s and FR2s
mentioned above.

The interesting question is: what physical phenomenon drives this rapid
fading of cluster AGN (and perhaps AGN in general)? 
Ellingson, Yee \& Green (1991) list two
possibilities: (1) the evolution of the cluster potential well greatly
reduces the number of interactions and mergers between gas rich galaxies
which could ``feed'' an AGN (e.g., Roos 1981); and (2) the rapidly
increasing ICM density prevents, in some as yet not understood way,
the ``feeding'' of the AGN (Stocke \& Perrenod, 1981). Both of these
hypotheses are supported by the direct X-ray observations of clusters
which show that: (1) in the richest clusters of galaxies a dense ICM is
already in place at $z\sim$1 (e.g., Donahue et al. 1999)
and (2) in poorer clusters at $z\sim$0.5
(wherein quasars
are found) there is no evidence for a dense ICM (e.g., Rector, Stocke \&
Ellingson 1995; Hall et al. 1997;
Harvanek et al. 2000),
but similar richness clusters at $z\sim$0 have dense ICMs as well as FR1
radio galaxies.
In fact, there is little
to distinguish these two hypotheses at low-z, since they both prevent
further feeding of the AGN as the ICM thickens and heats due to the
deepening of the potential well.

However, at the highest redshifts these two hypotheses predict opposite
outcomes. Extrapolating to $z\geq$4, before the epoch of quasars, 
the Roos (1981) interaction hypothesis would predict even greater
numbers of interactions (because the Universe was smaller and the number
of galaxies could only have been larger than today). Thus, by this
hypothesis, the highest observed quasars are the first AGN, whose
highest observed redshifts are due to supermassive black hole formation
timescales. However, by the Stocke \& Perrenod (1981) hypothesis, the
Universal ICM is denser at high-z, so that an AGN evolution opposite in
sense to what we observe at $z\leq$0.5 is possible; i.e., in this case,
the Black Holes are already formed at even higher redshifts as FR1s. 
Then these BL Lac/FR1s
brighten to become FR2s, which brighten to become quasars with cosmic
time. This idea was
first presented at the Como Conference on BL Lac Objects (Stocke 1987).
If this idea is correct, then very high-z BL Lac
Objects should be extremely numerous but hard to discover. Using
the observed SEDs of typical XBLs redshifted to $z\geq$4, the Sloan
digital sky survey (SDSS) would be the most sensitive search mechanism.
And, indeed, one strange AGN at $z$=4.7, which in many ways resembles a BL Lac
Object, has been found in the SDSS commissioning data (Fan et al. 2000)!
If this radical hypothesis for quasar evolution at high-$z$ is correct,
the SDSS should discover other examples of
this new class of high-z BL Lacs, which could contribute significantly
to the very faint X-ray source count (and to the X-ray Background).
Using the observed SEDs and luminosity functions of low-z BL Lacs, and
assuming that all quasars were once BL Lacs at higher redshifts
($z\geq$4), this new
class could contribute up to $\sim$1000 X-ray sources deg$^{-2}$ at
fluxes of 10$^{-15}$ ergs cm$^{-2}$ s$^{-1}$. And, if these objects have
SEDs similar to low-$z$ BL Lacs, they would be
spectrally very hard. 

\section{Acknowledgments}
JTS would like to thank Drs. R. Wurtz, E. Perlman \& T. Rector for their
past work, which contributes so importantly to this presentation, and 
their current interest in these strange objects. Anything of note and
veracity herein is due to them; all misconceptions are the exclusive
property
of their ex-thesis advisor.

\end{document}